\documentclass[RNAAS,twocolumn]{aastex701}
\usepackage{footnote}
\usepackage{graphicx} 
\usepackage{booktabs}   
\usepackage{array}      
\usepackage[utf8]{inputenc}
\usepackage{makecell}

\newcolumntype{L}{>{\centering\arraybackslash}p{2.8cm}}

\def\headline#1{%
\centerline{\hrulefill\quad #1 \quad\hrulefill}
}

\usepackage{hyperref}
\begin{document}

\title{A Spectral Atlas for Quiescent Galaxies}
\correspondingauthor{Massissilia L. Hamadouche}\email{mhamadouche@umass.edu}

\author[0000-0001-6763-5551]{Massissilia L. Hamadouche}
\affiliation{Department of Astronomy, University of Massachusetts, Amherst, MA 01003, USA}
\email{}
\author[0000-0002-0243-6575]{Jacqueline Antwi-Danso}\thanks{Dunlap Fellow}
\affiliation{David A. Dunlap Department of Astronomy \& Astrophysics, University of Toronto, 50 St George Street, Toronto, ON M5S 3H4, Canada}
\affiliation{Dunlap Institute for Astronomy \& Astrophysics, University of Toronto, 50 St George Street, Toronto, ON M5S 3H4, Canada}
\affiliation{Department of Astronomy, University of Massachusetts, Amherst, MA 01003, USA}
\email{}
\author[0000-0002-9861-4515]{Aliza Beverage}\thanks{NHFP Hubble Fellow}
\affiliation{Observatories of the Carnegie Institution for Science, 813 Santa Barbara Street, Pasadena, CA 91101, USA}
\affiliation{Department of Astrophysical Sciences, Princeton University, 4 Ivy Lane, Princeton, NJ 08544, USA}
\email{abeverage@carnegiescience.edu}
\author[0000-0001-7160-3632]{Katherine E. Whitaker}
\affiliation{Department of Astronomy, University of Massachusetts, Amherst, MA 01003, USA}
\affiliation{Cosmic Dawn Center (DAWN), Denmark}
\email{kwhitaker@astro.umass.edu}

\date{\today}

\begin{abstract}
      Key absorption features in the spectra of massive quiescent galaxies reveal a wealth of information about their stellar populations and, in some cases, the properties of gas within and around them. With the spectroscopic capabilities of \textit{JWST}, we are now able to probe deeper and farther into the near-infrared than ever before.  It is therefore crucial that we fully understand the origins of observed spectral absorption and emission features. The goal of this document is to collate important rest-frame optical to near-infrared (NIR) spectral features of quiescent galaxies in the context of their physical origin (e.g., stellar photospheres, the interstellar medium, or a combination thereof). We present a look-up table summarizing key information, including a ``diagnostic'' column indicating whether lines are most sensitive to age, metallicity, surface gravity (IMF sensitivity), $\alpha$-enhancement, electron temperature or gas density. This compilation is intended to serve as a practical reference for interpreting rest-optical-NIR spectral features in quiescent galaxies, particularly as \textit{JWST} spectroscopy enables increasingly detailed studies of their stellar populations and surrounding gas.
\end{abstract}

\section{Key features and Conventions}

Table~\ref{tab:absorption} lists important spectral features with rest-frame wavelengths ranging from 0.3 to 1.9~$\mu$m, separated into absorption and emission features. While some of these lines correspond to individual species, others represent blends of multiple transitions. Additional details are provided in the footnotes. All wavelengths are listed in vacuum. The vacuum wavelengths for most atomic lines are taken from the NIST Atomic Spectral Line Database. Molecular line wavelengths are taken from the referenced papers where possible.

In the following subsections, we discuss in more detail the commonly adopted Lick indices, the diagnostic utility of the hydrogen series (Section~\ref{hydrogen}), and spectral nuances associated with populations dominated by asymptotic giant branch (AGB) stars (Section \ref{agb}). 

\subsection{Lick indices}\label{lick}

Lick indices provide a standardized framework for measuring the strength of prominent absorption features in galaxy spectra and are widely used to constrain stellar population properties such as age, metallicity, and elemental abundance ratios. Although some Lick indices are associated with a dominant species, many measure blended absorption features arising from multiple atomic transitions or molecular bands. 

In Table \ref{tab:absorption}, we list the dominant absorption line (e.g. Fe {\scshape i} for Fe4383) instead of the Lick index. Fe indices from Fe4383 to Fe5782 are dominated by absorption mainly from Fe but with significant contributions from other iron-peak elements \citep{worthey1994,thomas_stellar_2003}. 
Ca4455 is a blend of Ca, Fe, Mn, and Ti \citep[to name only a few -- for more details see][]{worthey1994}. The dominant absorption for Ca4227 is due to Ca \textsc{i} ($\lambda_{\mathrm{rest, vac}} = 4227.92$\AA), and is therefore sensitive to $\alpha$-enhancement, as are the Mg\textsubscript{1} and Mg\textsubscript{2} indices (mainly measuring the Mg \textsc{i} absorption lines, see table note {\it g}). For more details on the primary abundances measured by the original Lick indices listed in Table \ref{tab:absorption}, we refer the reader to Section 2 and Table 1 in \cite{worthey1994}. 

\subsection{Balmer \& Paschen lines}\label{hydrogen}

At higher redshift, when the universe was younger and massive galaxies were rapidly transitioning from star-forming to quiescent systems, Balmer and Paschen absorption features become particularly important diagnostics of  star-formation and quenching timescales. Balmer lines observed in absorption are generally strongest in A-type stars though remain prominent in late-B through F-type stars. Paschen lines arise from the same hydrogen transitions but appear at longer wavelengths in the near-infrared and become increasingly visible in cooler stellar atmospheres (e.g., K and M-type stars). 
Lower-order Balmer or Paschen lines observed in emission in quiescent galaxies commonly signify the presence of an active galactic nucleus (AGN); an AGN can heat surrounding gas or drive highly ionized gas outflows. For clarity, we list higher-order Balmer absorption lines from H$\alpha$ up to H10 ($n=10$ to $n=2$ transition) as well as the Balmer limit, H$\infty$. The strengths of these lines in galaxy spectra correlate strongly with the ages of their stellar populations \citep{worthey1994}. Whereas the hydrogen absorption series varies only slightly once a galaxy ages beyond a few billion years \citep{delgado1999}, it produces the most prominent spectral variations at earlier times and therefore provides some of the most readily constrained spectral features at cosmic noon and earlier epochs.

\subsection{Molecular features from AGB stars}\label{agb}

Molecular absorption features associated with asymptotic giant branch (AGB) stars can strongly influence the near-infrared spectra of galaxies, particularly for intermediate-age stellar populations where thermally-pulsating (TP-) AGB stars may dominate the light \citep[e.g.,][]{maraston2006}. Because the evolution and spectra of these stars are uncertain, interpreting these features remains one of the more challenging aspects of stellar population modeling \citep[e.g.,][]{maraston2005}. 

In massive galaxies, carbon, nitrogen and oxygen mainly appear in molecular rather than atomic species, making it difficult to disentangle the effects of individual elemental abundances. Several Lick indices have been defined to help address this issue, for example the C$_2$4668 index to trace carbon abundance, and CN\textsubscript{1} or CN\textsubscript{2} to probe nitrogen abundance \citep[see also][]{gallazzi2005}. The deep CN edges observed in massive quiescent galaxy spectra \citep[as in][]{lu2025}), particularly at 1.1$\mu$m and 1.5$\mu$m, indicate the strong presence of carbon-rich TP-AGB stars.
Titanium oxide and vanadium oxide (VO) produce strong molecular absorption features in galaxy spectra and indicate the presence of cool M-type giant stars. 
At wavelengths longer than $\sim1\mu$m, VO and zirconium oxide (ZrO) become especially prominent in TP-AGB stars \citep{maraston2006,rayner2009,lu2025}. 
In contrast, CO and H$_2$O (water vapor) absorption features are commonly observed in oxygen-rich TP-AGB stars. These features provide useful diagnostics of stellar populations with ages between $\sim$0.4 and 2 Gyr \citep{maraston2005}. Together, these molecular signatures serve as tracers of intermediate-age stellar populations in galaxies and reflect the chemical enrichment from elements such as carbon, nitrogen (produced in AGB stars and CCSNe) and oxygen (only CCSNe), as well as heavier elements like Mn that also originate in Type Ia SNe. 

\section{Summary}

The rest-optical to near-infrared region of quiescent galaxy spectra is dominated by a range of features that trace their stellar populations and chemical enrichment.  These include hydrogen recombination series such as the Balmer and Paschen lines, strong metallicity indicators including magnesium, calcium, and iron, as well as molecular features associated with evolved stars (e.g., AGB stars) such as TiO, VO, and CN \citep{maraston2006,riffel2019,lu2025}. While hydrogen recombination lines are mostly observed in absorption in quiescent galaxies, they may also appear in emission in the presence of highly ionized gas. 

Elemental abundance patterns provide additional constraints on galaxy evolution.  $\alpha$ elements -- produced in massive stars ($>8 \mathrm{M_{\odot}}$) and released via core-collapse supernovae (such as Ca, O, Ne, Mg, Si, and S) -- and their abundances relative to iron-peak elements produced in Type Ia supernovae are widely used diagnostics of star formation histories in galaxies \citep[e.g.][]{thomas_stellar_2003}. 

The reference table presented in this work summarizes key spectral features across the rest-optical wavelength range and provides information on their physical and nucleosynthetic origins as well as diagnostic power. 
This compilation is intended to serve as a practical guide for interpreting the rest-frame optical to NIR spectra of quiescent galaxies, particularly in the era of high-sensitivity NIR spectroscopy enabled by facilities such as JWST.\\

\startlongtable
\begin{deluxetable*}{ccLLLLL} 
\setlength{\tabcolsep}{2.7pt}
\tablecolumns{7} 
\renewcommand{\arraystretch}{1.1}  
\tablecaption{Table of spectral absorption and emission features and their associated properties. Diagnostic column includes whether the feature can be used to trace age, effective temperature ($T_{\mathrm{eff}}$, stellar), electron temperature ($T_e$, gas), gas density, metallicity (Z/H), surface gravity (and hence, sensitivity to the IMF\tablenotemark{m}). The `source' column lists whether each feature is stellar, from the ISM or from AGN,  whereas the `line origin' column lists the physically observed origin (e.g. types of stars) and finally, the `nucleosynthetic channel' lists the dominant synthesis channel of the elements (including elements in molecules even if the molecule itself is formed in cool stellar atmospheres, e.g. CCSNe\tablenotemark{k}, SNeIa\tablenotemark{l}, AGB). For each feature, we list the values for the middle four columns in order of importance.}\label{tab:absorption}
\tablehead{
\colhead{\textbf{Feature}} & \colhead{$\lambda_{\mathrm{rest,vac}}$/{\AA}} &\colhead{\textbf{Source}} & \colhead{\textbf{Line origin}} & \colhead{\makecell[c]{\hspace{-32pt}\rule{0pt}{2.6ex}\textbf{Nucleosynthetic}\\[-2pt]\hspace{-32pt}\textbf{channel}}} & \colhead{\textbf{Diagnostic}} & \colhead{\textbf{References}}}
\startdata
\multicolumn{7}{c}{\textsc{\headline{Absorption Lines}}}\\
NH & 3360.00$^d$ & stellar & M-type stars & AGB, CCSNe  & Z/H, age & \scriptsize\cite{carbon1987,meyer1991} \phn\\
H$\infty$& 3645.98 & stellar &  A--G-type stars & BBN & age & \scriptsize\cite{wiese2009} \phn\\
\nodata & \nodata & \nodata & \nodata & \nodata & \nodata\\
H10 & 3798.98 & stellar & A--G-type stars & BBN & age & \scriptsize\cite{wiese2009}\phn\\
H9 & 3836.49 & stellar & A--G-type stars  & BBN & age & \scriptsize\cite{wiese2009} \phn \\
CN & 3884.10\tablenotemark{d} & stellar, ISM & late G/K giants & AGB, CCSNe & Z/H, age, $\alpha$/Fe & \scriptsize\cite{hesser1976,smolinksi2011,norris2013} \phn \\
H$\zeta$ & 3890.13 & stellar &  A--G-type stars & BBN & age & \scriptsize\cite{wiese2009} \phn \\
Ca \textsc{ii} K & 3934.78 & stellar, ISM\textsuperscript{i--iii} & F--M-type stars, neutral outflows & CCSNe, SNeIa & age, Z/H  & \scriptsize\cite{fraunhofer1817}  \phn \\
Ca \textsc{ii} H & 3969.59 & stellar & F--M-type stars\textsuperscript{i,iii}, neutral outflows & CCSNe, SNeIa & age, Z/H  & \scriptsize\cite{fraunhofer1817} \phn \\
H$\epsilon$ & 3971.20 & stellar & A--G-type stars & BBN & age  & \scriptsize\cite{wiese2009}\phn \\
Mn \textsc{i}\tablenotemark{j} & 4019.24 & stellar & late G/K stars & CCSNe, SNeIa & Z/H & \scriptsize\cite{kobayashi2006,kobayashi2020} \phn \\
Mn \textsc{i} & 4027.58 & stellar & late G/K stars & CCSNe, SNeIa & Z/H & \scriptsize\cite{kobayashi2006,kobayashi2020} \phn \\
Mn \textsc{i} & 4031.90 & stellar & late G/K stars & CCSNe, SNeIa & Z/H & \scriptsize\cite{kobayashi2006,kobayashi2020} \phn \\
H${\delta}$ & 4102.89 & stellar & A--G-type stars & BBN & age, Z/H, $\alpha$/Fe & \scriptsize\cite{worthey1994,wiese2009} \phn \\
SiH & 4143.00 & stellar, ISM  & late K--M stars & CCSNe, SNeIa & Z/H, age & \scriptsize\cite{davis1940,lambert1970,thomas2004} \\
CN$_1$\tablenotemark{f} & 4170.47 & stellar, ISM & G/K-type (C-rich) stars & AGB, CCSNe  & Z/H, age $\alpha$/Fe  & \scriptsize\cite{faber1985,thomas2003} \phn \\
CN$_2$\tablenotemark{f} & 4220.30 & stellar, ISM & G/K-type (C-rich) stars & AGB, CCSNe & Z/H, age, $\alpha$/Fe  & \scriptsize\cite{worthey1994,thomas2003} \phn \\
Sr \textsc{ii} & 4216.71 & stellar & late F--K-type stars & AGB & age & \scriptsize\cite{johnsonweinberg2020} \phn \\
Ca \textsc{i} & 4227.92 & stellar & late F--K type stars & CCSNe, SNeIa & Z/H, $\alpha$/Fe, age &\scriptsize\cite{worthey1994} \phn \\
Cr \textsc{i} & 4255.55 & stellar & late B--F type stars & SNeIa, CCSNe & Z/H, age & \scriptsize\cite{kobayashi2006} \phn \\
CH\tablenotemark{h} & 4305.61\tablenotemark{a} & stellar, ISM & late G--K-type (C-rich) stars & AGB, CCSNe & Z/H, age & \scriptsize\cite{faber1985,trager1998} \phn \\
H$\gamma$ & 4341.69 & stellar & A--G-type stars & BBN & age & \scriptsize\cite{wortheyottaviani1997,wiese2009} \phn \\
Fe \textsc{i} & 4384.77\tablenotemark{a} & stellar & F--K-type stars & SNeIa, CCSNe & Z/H, age & \scriptsize\cite{worthey1994} \phn \\
Ca \textsc{i} & 4464.24\tablenotemark{a} & stellar & late G-K-type stars & CCSNe, SNeIa & Z/H, $\alpha$/Fe, age & \scriptsize\cite{worthey1994} \phn \\
Fe \textsc{i} & 4537.76\tablenotemark{a} & stellar & F--K-type stars & SNeIa, CCSNe & Z/H, age & \scriptsize\cite{trager1998,yip2014} \phn \\
C$_2$\tablenotemark{b} & 4678.45 & stellar & late G--K (C-rich) stars & CCSNe, AGB & Z/H, age & \scriptsize\cite{trager1998,yip2014} \phn \\
H$\beta$ & 4862.68 & stellar & A--G-type stars & BBN & Z/H, age, $\alpha$/Fe & \scriptsize\cite{gonzalezdelgado1999,wiese2009}\phn \\
Fe \textsc{i} & 4977.75\tablenotemark{a} & stellar & F--K-type stars & SNeIa, CCSNe & Z/H, age & \scriptsize\cite{worthey1994}\phn \\
MgH & 5101.63 & stellar & F--K giants \& dwarfs & CCSNe & surface gravity, Z/H & \scriptsize\cite{grevesse1973,faber1985} \phn \\
Mg \textsc{i}\tablenotemark{g} & 5168.76 & stellar & F--M-type stars & CCSNe & $\alpha$/Fe, Z/H, age & \scriptsize\cite{deeming1960,faber1985} \phn \\
Mg \textsc{i} & 5174.13 & stellar & F--M-type stars & CCSNe & $\alpha$/Fe, Z/H, age &\scriptsize\cite{deeming1960,faber1985}\phn \\
MgH & 5175.38 & stellar & F--K giants \& dwarfs & CCSNe & surface gravity, Z/H  & \scriptsize\cite{grevesse1973,faber1985} \phn \\
Mg \textsc{i} & 5185.05 & stellar & F--M-type stars & CCSNe & $\alpha$/Fe, Z/H, age & \scriptsize\cite{deeming1960,faber1985} \phn \\
MgH & 5187.56 & stellar & F--K giants \& dwarfs & CCSNe & surface gravity & \scriptsize\cite{grevesse1973,faber1985} \phn \\
Fe \textsc{i} & 5265.65\tablenotemark{a} & stellar & F--K-type stars & SNeIa, CCSNe & Z/H, age, surface gravity & \scriptsize\cite{faber1985,worthey1994}\phn \\
Fe \textsc{i} & 5332.13\tablenotemark{a} & stellar & F--K-type stars & SNeIa, CCSNe & Z/H, age & \scriptsize\cite{worthey1994} \phn \\
Fe \textsc{i} & 5401.25\tablenotemark{a} & stellar & F--K-type stars & SNeIa, CCSNe& Z/H, age & \scriptsize\cite{worthey1994}  \phn \\
Fe \textsc{i} & 5710.25\tablenotemark{a} & stellar & F--K-type stars & SNeIa, CCSNe & Z/H, age & \scriptsize\cite{worthey1994}  \phn \\
Fe \textsc{i} & 5788.38\tablenotemark{a} & stellar & F--K-type stars & SNeIa, CCSNe & Z/H, age& \scriptsize\cite{worthey1994}  \phn \\
Fe \textsc{i} & 5848.60 & stellar & F--K-type stars & SNeIa, CCSNe & Z/H, age & \scriptsize\cite{worthey1994}  \phn \\
Fe \textsc{i} & 5902.47 & stellar & F--K-type stars & SNeIa, CCSNe & Z/H, age & \scriptsize\cite{worthey1994} \phn  \\
Na \textsc{i}\tablenotemark{e} & 5891.60  & stellar, ISM & late F--M-type stars, neutral outflows & CCSNe & Z/H, $N_{H_{\mathrm{i}}}$ & \scriptsize\cite{worthey1994,belli2024} \phn \\
Na \textsc{i} & 5897.60 & stellar, ISM & late F--M-type stars, neutral outflows & CCSNe & Z/H, $N_{H_{\mathrm{i}}}$  & \scriptsize\cite{worthey1994,belli2024} \phn \\
TiO$_1$ & 5967.13\tablenotemark{a} & stellar & M-type stars & CCSNe, SNeIa & Z/H, age, surface gravity  & \scriptsize\cite{faber1985,worthey1994,maraston2005,riffel2019}\phn \\
TiO$_2$ & 6232.63\tablenotemark{a} & stellar & M-type stars & CCSNe, SNeIa & Z/H, age, surface gravity & \scriptsize\cite{faber1985,worthey1994}\phn \\
Ca \textsc{i}\tablenotemark{c} & 6495.57 & stellar & late G-K-type stars & CCSNe, SNeIa & age, Z/H  & \scriptsize\cite{bica1986,schaerer2000,austin2007}\phn \\
H$\alpha$ & 6564.61 & stellar & A--G-type stars & BBN & age  & \scriptsize\cite{wiese2009}\phn \\
K \textsc{i} & 7667.01 & stellar, ISM & G--M-stars, neutral gas & CCSNe & Z/H  &
\scriptsize\cite{carretta2021}\\
K \textsc{i} & 7701.08 & stellar, ISM & G--M-stars, neutral gas & CCSNe & Z/H  &
\scriptsize\cite{carretta2021} \\
Na \textsc{i} & 8185.51\textsuperscript{viii} & stellar & M-type dwarfs & CCSNe & Z/H, age, surface gravity  &
\scriptsize\cite{delisle1992,schiavon1997,conroydokkum2012}\\
Na \textsc{i} & 8197.04\textsuperscript{viii} & stellar & M-type dwarfs & CCSNe & Z/H, age, surface gravity  &
\scriptsize\cite{delisle1992,schiavon1997,conroydokkum2012}\\
Pa20 & 8394.50 & stellar & K/M-type stars & BBN & age & \scriptsize\cite{wiese2009}\\
TiO\textsuperscript{v} & 8407.60 & stellar & M-type stars & CCSNe, SNeIa & Z/H, age & \scriptsize\cite{jorgenson1994,cenarro2009}\phn \\
Pa19 & 8415.42 & stellar & K/M-type stars & BBN & age & \scriptsize\cite{wiese2009}\\ 
Pa18 & 8440.07 & stellar & K/M-type stars & BBN & age & \scriptsize\cite{wiese2009}\\ 
Pa17 & 8469.37 & stellar & K/M-type stars & BBN & age & \scriptsize\cite{wiese2009} \phn\\ 
Ca \textsc{ii}& 8500.36 & stellar & late F--M-type stars & CCSNe, SNeIa & age, Z/H, surface gravity  & \scriptsize\cite{barbuy1989,bosler2002,mcconnell2016}\phn \\
Pa16 & 8504.61 & stellar & K/M-type stars & BBN & age & \scriptsize\cite{wiese2009}\\ 
Pa15 & 8547.52 & stellar & K/M-type stars & BBN & age &\scriptsize\cite{wiese2009} \\ 
Ca \textsc{ii} & 8544.44 & stellar & late F--M-type stars & CCSNe, SNeIa  & age, Z/H, surface gravity   & \scriptsize\cite{barbuy1989,bosler2002,mcconnell2016}\phn \\
Pa14 & 8600.54 & stellar & K/M-type stars & BBN & age &\scriptsize\cite{wiese2009} \\ 
Ca \textsc{ii} & 8664.52 & stellar & late F--M-type stars & CCSNe, SNeIa & age, Z/H, surface gravity & \scriptsize\cite{barbuy1989,bosler2002,mcconnell2016}\phn \\
Pa13 & 8667.18 & stellar & K/M-type stars & BBN & age & \scriptsize\cite{wiese2009}\\ 
Fe \textsc{i} & 8691.01 & stellar & late G--K-type stars & SNeIa, CCSNe & Z/H  & \scriptsize\cite{worthey1994}\\ 
Pa12 & 8752.65 & stellar & K/M-type stars & BBN & age & \scriptsize\cite{wiese2009}\\ 
Mg \textsc{i} & 8809.18 & stellar & F--M-type stars & CCSNe & age, Z/H, $T_{\mathrm{eff}}$, surface gravity  & \scriptsize\cite{mcconnell2016} \\
Pa11 & 8864.99 & stellar & K/M-type stars & BBN & age & \scriptsize\cite{poetker1927,wiese2009}\\ 
TiO\textsuperscript{v} & 8870.90 & stellar & M-type stars & CCSNe & Z/H, $T_{\mathrm{eff}}$, surface gravity  & \scriptsize\cite{jorgenson1994,maraston2005,cenarro2009}\phn \\
Pa10 & 9017.15 & stellar & K/M-type stars & BBN & age & \scriptsize\cite{poetker1927,wiese2009}\\ 
Pa9 & 9231.30 & stellar & K/M-type stars & BBN & age & \scriptsize\cite{poetker1927,wiese2009}\\ 
CN\textsuperscript{vi} & 9320.00 & stellar & C-rich (C-type) stars & AGB, CCSNe & Z/H, age & \scriptsize\cite{riffel2015,lu2025}\phn \\
Pa$\epsilon$ & 9548.32 & stellar & K/M-type stars & BBN & age & \scriptsize\cite{poetker1927,wiese2009}\phn \\
FeH & 9915.90\textsuperscript{iv} & stellar & late M-dwarfs & SNeIa, CCSNe & surface gravity &\scriptsize\cite{mcconnell2016}\\
FeH  & 10036.47\textsuperscript{iv} & stellar & late M-dwarfs & SNeIa, CCSNe & surface gravity & \scriptsize\cite{schiavon1997}\phn \\
Pa$\delta$ & 10052.10 & stellar & K/M-type stars & BBN & age & 
\scriptsize\cite{poetker1927,wiese2009}\phn\\
VO & 10480.00 & stellar & late M-type stars & CCSNe, SNeIa & Z/H, $T_{\mathrm{eff}}$, surface gravity & \scriptsize\cite{mckellar1956,spinrad1966,riffel2015}\phn \\
Pa$\gamma$ & 10941.20 & stellar & K/M-type stars & BBN & age & \scriptsize\cite{poetker1927,wiese2009}\phn \\
TiO & 11044.80 & stellar & M-type stars & CCSNe, SNeIa & age, surface gravity  & \scriptsize\cite{jorgenson1994,cenarro2009}\phn \\
CN & 11100.00 & stellar & C-rich (C-type) stars & AGB, CCSNe & age, Z/H  & \scriptsize\cite{maraston2005,riffel2015}\phn \\
Na \textsc{i} & 11384.57 & stellar & M-dwarfs & CCSNe & surface gravity & \scriptsize\cite{eftekhari2021}\phn \\
Fe \textsc{i} & 11384.57 & stellar & late G/K-type stars & SNeIa, CCSNe & Z/H  & \scriptsize\cite{eftekhari2021}\phn \\
Pa$\beta$ & 12821.60 & stellar & K/M-type stars & BBN & age & \scriptsize\cite{poetker1927,wiese2009,eftekhari2021}\phn\\
Al \textsc{i} & 13154.35 & stellar & K/M-type stars & CCSNe, AGB & Z/H  & \scriptsize\cite{kobayashi2006,roeck2015,nordlander2017,eftekhari2021}\phn \\
H$_2$O & 14000.00 & stellar & O-rich M-type stars & CCSNe & age & \scriptsize\cite{riffel2015}\phn\\
CN & 14175.00 & stellar & C-rich (C-type) stars & AGB, CCSNe & age, Z/H  & \scriptsize\cite{maraston2005,riffel2015}\phn \\
CN & 14607.50 & stellar & C-rich (C-type) stars & AGB, CCSNe & age, Z/H  & \scriptsize\cite{maraston2005,riffel2015}\phn \\
Mg \textsc{i} & 14881.60 & stellar & F--M type stars & CCSNe & Z/H  & \scriptsize\cite{rayner2009}\phn \\
CO & 15772.50 & stellar & C/O-rich (C/M-type) stars & AGB, CCSNe & age & \scriptsize\cite{riffel2015}\\
CO & 16230.00 & stellar & C/O-rich (C/M-type) stars & AGB, CCSNe & age & \scriptsize\cite{riffel2015}\\
H$_2$O & 17100.00 & stellar & O-rich (M-type) stars & CCSNe & age & \scriptsize\cite{riffel2015}\phn\\
CN & 17700.00 & stellar & C-rich (C-type) stars & AGB, CCSNe & age, Z/H  & \scriptsize\cite{lu2025}\\
C$_{2}$ & 17800.00 & stellar & C-rich (C-type) stars & AGB, CCSNe & age, Z/H & \scriptsize\cite{maraston2005,lu2025}\\
Pa$\alpha$ & 18756.10 & stellar & K/M-type stars & BBN & age & \scriptsize\cite{poetker1927,wiese2009}\phn\\
H$_2$O & 19000.00 & stellar & O-rich (M-type) stars & CCSNe & age & \scriptsize\cite{riffel2015}\phn\\
\multicolumn{7}{c}{\textsc{\headline{Emission Lines}}}\\
{[Ne \textsc{iii}]}\tablenotemark{i} & 3342.18 & AGN & NLR, post-AGB, shocks & CCSNe & high-ionization & \scriptsize\cite{brotherton1999}\\
{[Ne \textsc{v}]}\tablenotemark{i} & 3346.79 & AGN & CLR\footnote{CLR = Coronal Line Region, NLR = Narrow Line Region, BLR = Broad Line Region} & CCSNe &  high-ionization & \scriptsize\cite{baldwin1981}\\
{[Ne \textsc{v}]} & 3426.85 & AGN, ISM & CLR, shocks & CCSNe & high-ionization & \scriptsize\cite{baldwin1981}\phn \\
{[O \textsc{ii}]} & 3727.10 & AGN, ISM & NLR, outflows & CCSNe & $T_{e}$, gas density, SFR & \scriptsize\cite{osterbrock1990}\phn \\
{[O \textsc{ii}]} & 3729.86 & AGN, ISM & NLR, outflows & CCSNe & $T_{e}$, gas density, SFR  & \scriptsize\cite{osterbrock1990}\phn \\
{[Ne \textsc{iii}]}\tablenotemark{i} & 3869.86 & AGN, ISM & NLR, post-AGB, shocks & CCSNe & gas density & \scriptsize\cite{belli2024}\phn \\
He \textsc{ii} & 4687.02 & AGN, stellar & AGN, WR winds & BBN, CCSNe & gas density, $T_{\mathrm{eff}}$  & \scriptsize\cite{osterbrock1990,worthey1994}\phn\\
{[Fe \textsc{vii}]}\tablenotemark{i} & 4894.80 & AGN & NLR & SNeIa, CCSNe & high-ionization & \scriptsize\cite{osterbrock1990}\phn \\
{[O \textsc{iii}]} & 4960.29 & AGN, ISM & NLR, outflows & CCSNe & $T_{e}$, gas density & \scriptsize\cite{nussbaumer1981}\phn \\
{[O \textsc{iii}]} & 5008.24 & AGN, ISM & NLR, outflows & CCSNe & $T_{e}$, gas density & \scriptsize\cite{nussbaumer1981}\phn \\
He \textsc{i} & 5877.25 & AGN, stellar & BLR, WR winds & BBN, CCSNe & surface gravity, $T_{\mathrm{eff}}$, high-ionization & \scriptsize\cite{martin1997,schaerer2000}\phn \\
{[N \textsc{ii}]} & 6549.86 & AGN, stellar & NLR, post-AGB shocks & AGB, CCSNe & $T_{e}$, Z/H & \scriptsize\cite{baldwin1981}\phn \\
H$\alpha$ & 6564.61 & ISM, AGN & BLR/NLR, H\textsc{ii} regions & BBN & $M_{\mathrm{BH}}$,  SFR & \scriptsize\cite{wiese2009}\phn \\
{[N \textsc{ii}]} & 6585.27 & AGN, stellar & NLR, post-AGB, shocks & AGB, CCSNe&  $T_{e}$, Z/H & \scriptsize\cite{kirhakos1989}\phn \\
{[S \textsc{ii}]} & 6718.29  & AGN & NLR, post-AGB, shocks  & CCSNe & $T_{e}$, ionization (with [S\textsc{iii}]), gas density & \scriptsize\cite{kirhakos1989}\phn \\
{[S \textsc{ii}]} & 6732.68 & AGN, stellar & NLR, post-AGB, shocks & CCSNe & $T_{e}$, ionization (with [S\textsc{iii}]), gas density & \scriptsize\cite{baldwin1981} \phn\\
{[Fe \textsc{vii}]}\tablenotemark{i} & 6088.61 & AGN & CLR & SNeIa, CCSNe & $T_{e}$, gas density & \scriptsize\cite{osterbrock1982}\phn \\
{[Ar \textsc{iii}]} & 7137.77 & AGN & NLR & CCSNe & high-ionization & \scriptsize\cite{schaerer2000} \phn\\
{[S \textsc{iii}]} & 9071.32 & AGN, ISM & NLR & CCSNe & $T_{e}$, Z/H, (ionization with [S\textsc{ii}])  &  \scriptsize\cite{kirhakos1989} \phn\\
{[S \textsc{iii}]} & 9533.20 & AGN, ISM & NLR & CCSNe & $T_{e}$, Z/H, (ionization with [S\textsc{ii}])  & \scriptsize\cite{kirhakos1989} \phn\\
He \textsc{i} & 10031.20 & ISM & NLR/BLR & BBN, CCSNe & $T_{e}$, gas density, $T_{\mathrm{eff}}$ & \scriptsize\cite{osterbrock1990}\phn\\
He \textsc{i} & 10830.00\textsuperscript{vii} & AGN, ISM, stellar & NLR/BLR, WR winds & BBN, CCSNe & $T_{e}$, gas density, $T_{\mathrm{eff}}$  & \scriptsize\cite{osterbrock1990,worthey1994}\\
\enddata 
\tablenotetext{a}{These are Lick indices measuring the abundance of the main absorber (e.g. Ti for TiO\textsubscript{1,2} and Fe \textsc{i} for the iron Lick indices Fe4383, Fe4531, Fe5015, Fe5270, Fe5335, Fe5406, Fe5709, Fe5782) within the central band-pass. We have given the central wavelength of this band-pass as the index wavelength. Finally, Fe is produced almost equally in abundance via both CCSNe and SNeIa.}
\tablenotetext{b}{This was previously known as the Fe4668 index, but was found to be more sensitive to changes in carbon abundance than iron, and was thus renamed C$_2$4668.}
\tablenotetext{c}{This absorption feature is a blend of Ti \textsc{i} at 6499.48 {\AA}, Ba \textsc{i} at 6500.56  {\AA}, Fe \textsc{i} at 6500.80 {\AA}, Ca \textsc{i} at 6501.45 {\AA}, and Fe \textsc{i} at 6503.48 {\AA} \citep[see also][]{austin2007}.}
\tablenotetext{d}{This is a broad molecular feature ranging from $\lambda_{\mathrm{min}}$ to $\lambda_{\mathrm{max}}$. Due to the molecular lines falling close together (due to different rotational/vibrational transitions) and becoming unresolvable, this causes a region of maximum absorption, coined a `bandhead'.}
\tablenotetext{e}{Na I is a doublet corresponding to the fine-structure splitting of Na I (neutral sodium) excited states at 5889 {\AA} and 5896 {\AA}. The Na I doublet is insensitive to $\alpha$/Fe enhancements, and is thought to include the contribution from cool stars (e.g. G-types and cooler), with excess absorption in the spectra of massive galaxies attributed to ionized outflows from the interstellar medium \citep{thomas2003,baron2022}. Recent observations have attributed strong (blue- and red-shifted) absorption to multiphase outflows in quiescent galaxies \citep[][]{belli2024}.} 
\tablenotetext{f}{CN$_2$ is the newer Lick index definition of CN$_1$, where the only difference from CN$_1$ is in the blue pseudo-continuum range which avoids the H$\delta$ feature. \textit{Blue:} 4085.125-4097.625 {\AA} (CN$_1$: 4081.375-4097.625 {\AA}), \textit{Red:} 4245.375-4285.375 {\AA}, and the index region is 4143.375-4178.375 {\AA}. CN$_2$ is more appropriate when measuring stellar populations with young, hot stars.}
\tablenotetext{g}{Here, we only list the lines in the Mg\textit{b} complex; the three Mg \textsc{i} absorption lines. The Mg$_1$ (5071.00-5134.75 {\AA}) and Mg$_2$ (5156-5197 {\AA}) Lick indices measure mostly MgH, and MgH $+$ Mg$b$, respectively \citep[with a small contribution for both indices from Fe \textsc{i}][]{worthey1994}. Mg$b$ measures the equivalent width of the combined absorption from the Mg \textsc{i} triplet. Mg$b$ is also sensitive to surface gravity \citep{faber1985}.}
\tablenotetext{h}{The Fraunhofer \citep[][]{fraunhofer1817} G line (G4300 index) is actually a blend of absorption lines that includes CH, Fe and Ca.}
\tablenotetext{i}{[Ne V], [Ne III] are produced in high-ionization regions, mostly by active galactic nuclei, but also in regions of intense star-formation (e.g. starbursts).}
\tablenotetext{j}{Mn I is mainly produced in core-collapse supernovae, in the outer layers of massive stars during explosive, incomplete silicon burning where $^{55}$Co decays to $^{55}$Mn \citep{barbuy2013}. A small fraction of Mn I is also produced in Type Ia supernovae \citep{kobayashi2006} and more Mn is produced in Type Ia supernovae than iron, so is important in constraining the physics of Type Ia SNe \citep[see also][]{kobayashi2006}.}
\tablenotetext{k}{For most of the elements listed as CCSNe, these are actually formed in the core burning of the progenitor massive stars ($> 8 M_{\odot}$) then released into the ISM via CCSNe.}
\tablenotetext{l}{Type Ia supernovae are thought to arise from the thermonuclear explosions of carbon-oxygen white dwarf (WD) stars in binary systems, caused by either the accretion of material from the companion star, or from a merger with a secondary WD. Type II supernovae arise from the core collapse of massive red supergiant (RSG) stars.}
\tablenotetext{m}{Strong surface gravity features are a tracer of the ratio of dwarf to giant stars in a stellar population. For example, the calcium triplet (CaT) is sensitive to changes in surface gravity; the absorption strength increases in strength with \textit{decreasing} surface gravity, making this a \textit{giant-sensitive} indicator and thus sensitive to the IMF of the stellar population.}
\tablecomments{\\
\textsuperscript{(i)}Photospheric lines are strong in late-type stars (F, G, K, M stars) due to low ionization energies. Ca II is also more abundant in cooler stars, where calcium remains partially ionized. The K line is generally stronger than the H line because calcium is predominantly singly ionized in the cooler stellar atmospheres where these lines form.
\textsuperscript{(ii)}Ca K can also arise in diffuse neutral or warm ionized ISM gas, and is often accompanied by excess Na I D absorption in ISM studies.  Ca H is often weaker or absent in the ISM as it is not easily formed or stable in diffuse or dense ISM.
\textsuperscript{(iii)}Ca is primarily produced in massive stars ($>$8 M$_{\odot}$) through SNe nucleosynthesis, thus primarily Type II SNe, but Type Ia SNe can produce Ca via Si burning. \phn 
\textsuperscript{(iv)}The FeH Wing-Ford band ranges from 9850–10200 \AA\ and contains contributions from FeH, Fe I, CN and Ca II. The WFB, as with FeH, is a very sensitive surface gravity indicator, and this feature is stronger in dwarf stars. CN dominates in giants.
\textsuperscript{(v)}In galaxies, TiO has been found to reflect the overall metallicity of a stellar population. In stars, TiO is a good discriminator between giants and dwarfs in M-type stars \citep[see e.g.][]{cenarro2009}.
\textsuperscript{(vi)}The CN feature here is a combination of absorption from TiO, ZrO and CN.
\textsuperscript{(vii)}This is the near-infrared He I triplet. Lines are at 10832.1, 10833.2, and 10833.3 {\AA}.
\textsuperscript{(viii)}Na I at 0.82$\mu$m is a blend of the Na I $\lambda$8183, $\lambda$8195 doublet and the TiO band-head at a resolution of 300 km s$^{-1}$ \citep[see][]{schiavon1997}. Na I at 1.14 microns is also affected by molecular H$_2$O which can weaken the sodium absorption in later type stars (L-dwarfs). 
}
\end{deluxetable*}


\clearpage

\section*{Acknowledgments}
The Dunlap Institute is funded through an endowment established by the David Dunlap family and the University of Toronto. MLH would also like to thank S. Bodansky for helpful comments. 
\bibliographystyle{aasjournalv7}
\bibliography{refs}

\end{document}